\documentstyle[epsfig]{aipproc}

\begin{document}
\title{Luminosities, Space Densities\\
and Redshift Distributions\\
of Gamma-Ray Bursts}

\author{Maarten Schmidt}
\address{California Institute of Technology, Pasadena, California 91125}

\maketitle

\begin{abstract}
We use the BD2 sample of gamma-ray bursts (GRBs) based on 5.9 years
of BATSE DISCLA data with a variety of models of the luminosity function 
to derive characteristic GRB luminosities, space densities and redshift 
distributions. Previously published results for an open universe and
modest density evolution of the GRBs showed characteristic peak
luminosities around 
$5 \times 10^{51}$ ergs s$^{-1}$ in the $50-300$ keV band if the emission 
is isotropic, and local space densities around 0.2 Gpc$^{-3}$ y$^{-1}$.
In this paper, we illustrate for several luminosity function
models the predicted distributions of peak flux, luminosity and 
redshifts. We use the luminosity function models also to address the 
connection between supernovae and GRB. If all supernovae of type Ib/c
harbor a GRB, the beaming fraction would have to be in the range 
$10^{-5} - 10^{-3.5}$. We find that GRB 980425, if correctly identified 
with SN 1998bw, has to be part of a population different from that of the 
bulk of GRBs.  

\end{abstract}

\begin{figure}[t!]
\centerline{\epsfig{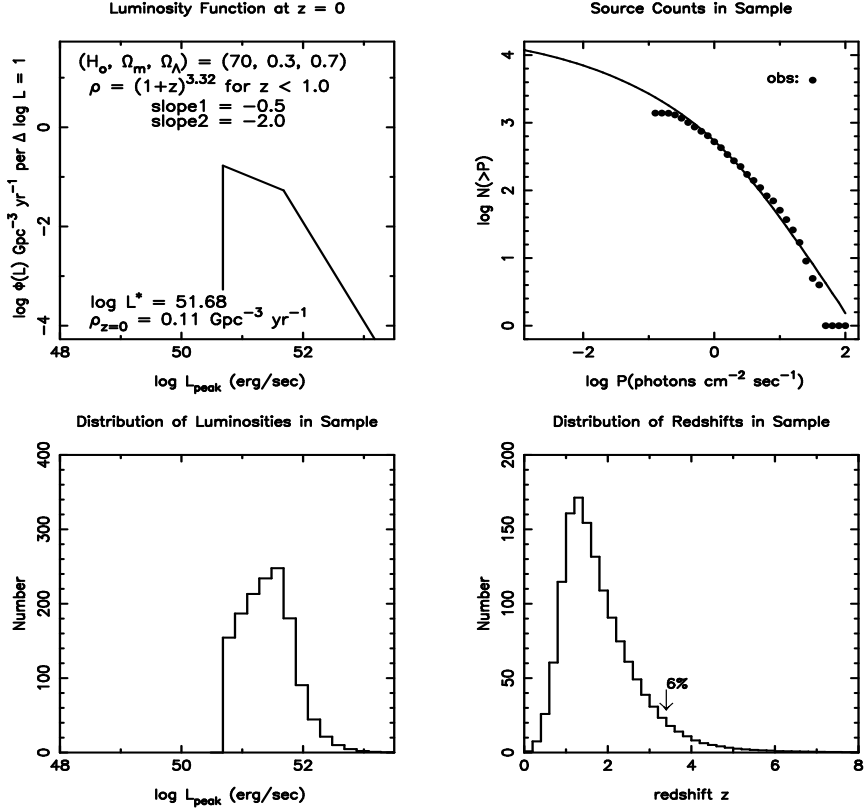}}
\vspace{10pt}
\caption{Assuming a luminosity function extending from $L^*/10$ to
$100L^*$ characterized in the upper left panel and the text, the other
panels show the predicted luminosity distribution, source counts and
redshift distributions for the BD2 sample of GRBs}
\label{F:mschmidt:1}
\end{figure}

\section*{Introduction}
The luminosity function and cosmological evolution of extragalactic objects
are usually derived from observed samples that are complete above a given
flux limit and have measured redshifts. In the case of gamma-ray bursts
(GRBs), there is no well defined complete sample with redshifts available at 
the present time. Under these circumstances, we have found it useful to 
invert the process, to {\it assume} a luminosity function, and to derive 
intrinsic properties of GRBs such as their characteristic peak luminosity 
$L^*$ and local space density \cite{schmidt99a}. In this paper, we illustrate
the predictions for some of these luminosity function models. 
We also use these models to address the issue of a connection between 
supernovae and GRBs.

\section*{Models of the GRB Luminosity Function}

\begin{figure}[t!]
\centerline{\epsfig{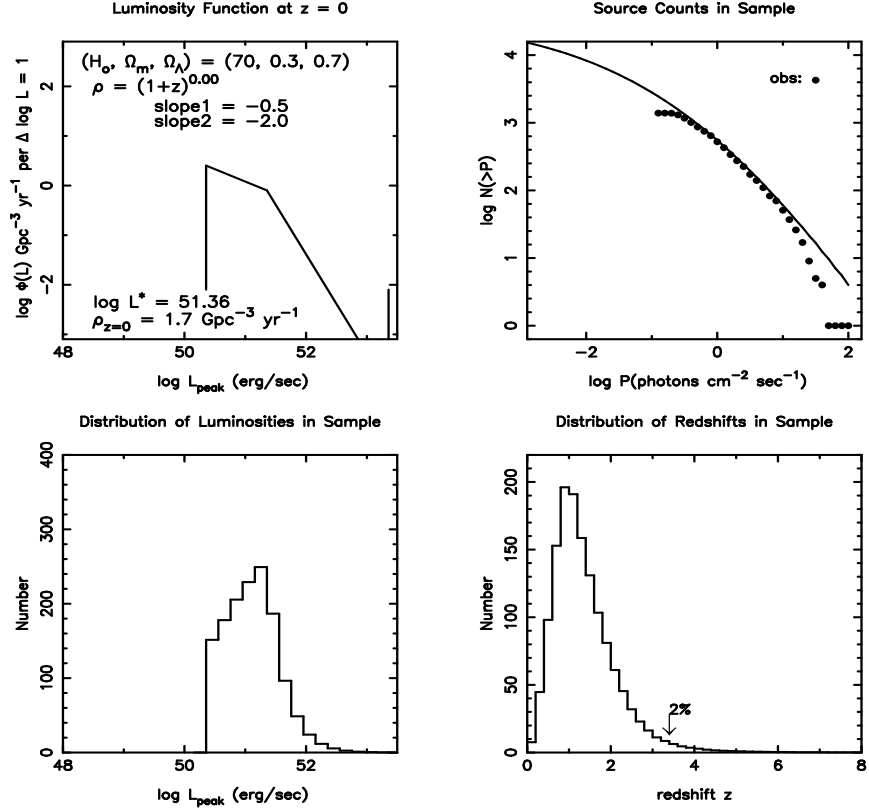}}
\vspace{10pt}
\caption{Predictions based on a GRB luminosity function with no density
evolution}
\label{F:mschmidt:2}
\end{figure}

We start with a brief description of the methodology. For a detailed 
description of the models, the reader is referred to \cite{schmidt99a}. 
We used luminosity functions that were broken power laws with characteristic
luminosity $L^*$, with different slopes and different extents above and
below $L^*$. In addition, we assumed density evolution rising to a factor of
10 at $z=1$ and constant for $z>1$. We used the BD2 sample of 1391 GRB which
was derived from 5.9 years of BATSE DISCLA data on a timescale of 1024 msec
\cite{schmidt99a,schmidt99b}. For a given luminosity function model, 
the observed euclidean value of $V/V_{max} = 0.334 \pm 0.008$ allowed a robust
determination of $L^*$. The total number of GRBs in the BD2 sample provided
the normalization of the luminosity function.
For a given cosmology and evolution, the value of $L^*$ ranged
over a factor of 6 and the local space density over a factor of 2.
Typical values\footnote{These values are based on fluxes corrected for 
a scale error, mentioned in a footnote in \cite{schmidt99a}.}
for an open universe with $q_o = 0.1$ were a characteristic peak luminosity
$L^* = 5 \times 10^{51}$ ergs s$^{-1}$
and a local space density around 0.2 Gpc$^{-3}$ yr$^{-1}$. Beaming 
reduces the peak luminosity and increases the density by the same factor.

We illustrate in this paper the results based on several different
luminosity functions. We are using
a cosmological model that is a flat accelerating universe with a matter 
density $\Omega_m = 0.3$ and cosmological constant $\Omega_{\Lambda} = 
0.7$ \cite{bahcall99}. We assume a Band type GRB spectrum \cite{band93}
with $\alpha = -1$, $\beta = -2$, and break energy $E_o = 150$ keV.
In Figure \ref{F:mschmidt:1}, we show the results 
for a typical model of the luminosity function, extending from $L^*/10$
to $100L^*$. The density evolution $\rho = (1+z)^{3.32}$ amounts to a
factor of 10 at $z=1$ and is then constant at $z>1$. The local density
is 30\% lower than that for an open universe with $q_o = 0.1$
\cite{schmidt99a}. As shown in
Figure \ref{F:mschmidt:1}, the predicted $N(>P)$ distribution agrees
well with the observations. The largest GRB redshift so far observed is 3.42 
for GRB 971214 \cite{kulkarni98}. Based on the expected redshift distribution,
the probability of finding a redshift of $z=3.4$ or larger in the BD2
sample is 6\%. The median redshift is 1.5 and the largest redshift 
among the 1391 GRB in the BD2 sample is $z=9.3$ according to this model.
 
\begin{figure}[t!]
\centerline{\epsfig{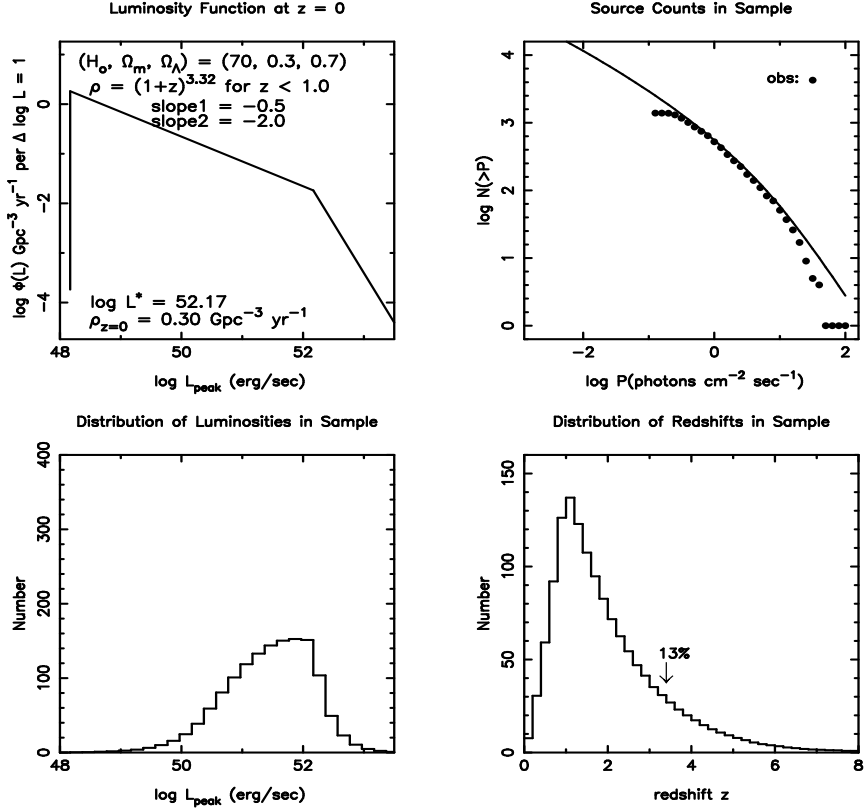}}
\vspace{10pt}
\caption{Predictions based on a luminosity function extending from
$L^*/10000$ to $100L^*$}
\label{F:mschmidt:3}
\end{figure}

In Figure \ref{F:mschmidt:2}, we show the results if there is no density
evolution. Compared to the previous case, $L^*$ is down by a factor of 2
and the local space density up by a factor of 15. The predicted source 
counts fall below the observed points. The redshift distribution has 
only 2\% of GRBs at a redshift of 3.4 or above.

\section*{The Supernova Connection}

If the light curves of GRB afterglows show evidence for a supernova of 
Type Ib/c \cite{bloom99}, the question arises whether all such supernovae
could be associated with GRBs. In the following, we update a discussion of
this issue by Lamb\cite{lamb99}. If the gamma-ray emission is beamed over a 
fraction $f_{beam}$ of the celestial sphere, the local GRB rate is roughly
(0.1 - 3) $f_{beam}^{-1}$ Gpc$^{-3}$ yr$^{-1}$, where the range mostly reflects
the effect of density evolution (up to a factor of 10 at $z=1$), or no density
evolution \cite{schmidt99a}.
The rate of Type Ib/c supernovae \cite{cap97} in spiral galaxies is 
$\sim 10^{-13}$ L$_{\sun}^{-1}$ yr$^{-1}$. With a luminosity density
of $\sim 10^8$ L$_{\sun}$ Mpc$^{-3}$ \cite{san88} for spirals, this
corresponds to $\sim 10^4$ SN Gpc$^{-3}$ yr$^{-1}$. 
If $f_{beam} = 10^{-2}$, then only 1 in (1000-30) SN Type Ib/c could have an 
associated GRB. If every Type Ib/c supernova harbors a GRB, $f_{beam}$ would 
have to be as small as $10^{-5} - 10^{-3.5}$. 

\begin{figure}[t!]
\centerline{\epsfig{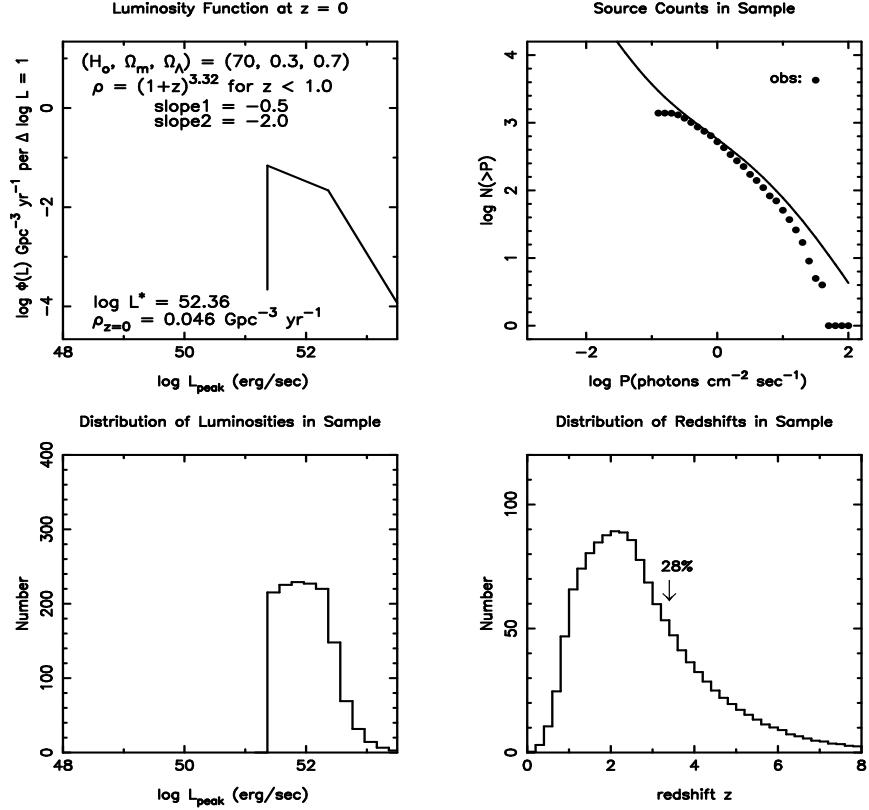}}
\vspace{10pt}
\caption{Assuming that 30\% of observed GRB are of low luminosity, 
we show predicted distributions for the remaining GRBs for a given
model of the luminosity function.}
\label{F:mschmidt:4}
\end{figure}

If GRB 980425 is associated with the SN 1998bw , its peak luminosity 
is $\log L = 46.7$ \cite{galama99}. We explored a luminosity function 
model that extends a factor of $10^4$ below $L^*$ (Fig. \ref{F:mschmidt:3}). 
Even though the luminosity function reaches down
to around $\log L \sim 48$, the distribution of {\it observed}
luminosities predicts only one GRB to have 
$\log L < 48.5$, or a probability of around 0.1\%. 
Given this low probability, the identification of GRB 980425 with 
SN 1998bw can only be understood if it represents a separate population 
of low-luminosity bursts. Assuming that 30\% of the observed GRBs are 
of low luminosity, we show in Figure \ref{F:mschmidt:4} the expected 
distributions for the remaining GRBs assuming that they have a luminosity 
function identical in shape to that of Figure \ref{F:mschmidt:1}. The fit of 
$N(>P)$ to the observed distribution is poor, indicating that at most 30\% 
of observed GRBs can have a low luminosity.

\end{document}